\documentstyle[12pt,fleqn,epsf]{article}

\catcode`@=11
\def\chkspace{%
  \relax   
  \begingroup\ifhmode\aftergroup\dochksp@ce\fi\endgroup}
\def\dochksp@ce{%
  \unskip              
  \futurelet\chkspct@k\d@chkspc  
}
\def\d@chkspc{%
  \let\nxtsp@ce=\relax
  \ifx\chkspct@k.\else     
    \ifx\chkspct@k,\else
      \ifx\chkspct@k;\else
        \ifx\chkspct@k!\else
          \ifx\chkspct@k?\else
            \ifx\chkspct@k:\else
              \ifx\chkspct@k)\else
              \ifx\chkspct@k(\else
                \ifx\chkspct@k]\else
                  \ifx\chkspct@k-\else
                    \ifx\chkspct@k\egroup\else  
                      \let\nxtsp@ce=\put@space  
                    \fi
                  \fi
                \fi
              \fi
              \fi
            \fi
          \fi
        \fi
      \fi
    \fi
  \fi
  \nxtsp@ce
}
\def\put@space{$\;$}
\catcode`@=12

\def\ra{{$\rightarrow$}\chkspace}

\def\etal{{\it et al.}\chkspace}

\def\adhoc{{\it ad hoc}\chkspace}

\def\gluino{\relax\ifmmode \tilde{g} \else $\tilde{g}$ \fi\chkspace}

\chkspace
\chkspace
\def\m0{$M_{0}$}\chkspace
\def\m0m{$M_{0}MAX$}\chkspace
\chkspace
\chkspace

\def\bbrm{\relax\ifmmode {\rm b}\bar{\rm b}
       \else ${\rm b}\bar{\rm b}$ \fi\chkspace}

\def\ccrm{\relax\ifmmode {\rm c}\bar{\rm c}
       \else ${\rm c}\bar{\rm c}$ \fi\chkspace}

\def\tt{\relax\ifmmode {\rm t}\bar{\rm t}
       \else ${\rm t}\bar{\rm t}$ \fi\chkspace}
\def\ss{\relax\ifmmode {\rm s}\bar{\rm s}
       \else ${\rm s}\bar{\rm s}$ \fi\chkspace}
\def\uu{\relax\ifmmode {\rm u}\bar{\rm u}
       \else ${\rm u}\bar{\rm u}$ \fi\chkspace}
\def\dd{\relax\ifmmode {\rm d}\bar{\rm d}
       \else ${\rm d}\bar{\rm d}$ \fi\chkspace}

\def\qqg{\relax\ifmmode {\rm q}\bar{\rm q}{\rm g}
\else q$\bar{\rm q}$g \fi\chkspace}

\def\afb{\relax\ifmmode A_{FB} \else
{{$A_{FB}$}}\fi\chkspace}
\def\afbb{\relax\ifmmode A_{FB}^b \else
{{$A_{FB}^b$}}\fi\chkspace}
\def\pafb{\relax\ifmmode \tilde{A}_{FB} \else
{{$\tilde{A}_{FB}$}}\fi\chkspace}
\def\pafbb{\relax\ifmmode \tilde{A}_{FB}^b \else
{{$\tilde{A}_{FB}^b$}}\fi\chkspace}

\def\pafbzo{\relax\ifmmode \tilde{A}_{FB}|_{O(0)} \else
{{$\tilde{A}_{FB}|_{O(0)}$}}\fi\chkspace}
\def\pafbfo{\relax\ifmmode \tilde{A}_{FB}|_{\oalp} \else
{{$\tilde{A}_{FB}|_{\oalp}$}}\fi\chkspace}
\def\pafbso{\relax\ifmmode \tilde{A}_{FB}|_{\oalpsq} \else
{{$\tilde{A}_{FB}|_{\oalpsq}$}}\fi\chkspace}
\def\pafbto{\relax\ifmmode \tilde{A}_{FB}|_{\oalpc} \else
{{$\tilde{A}_{FB}|_{\oalpc}$}}\fi\chkspace}

\def\pafbbzo{\relax\ifmmode \tilde{A}_{FB}^b|_{O(0)} \else
{{$\tilde{A}_{FB}^b|_{O(0)}$}}\fi\chkspace}
\def\pafbbfo{\relax\ifmmode \tilde{A}_{FB}^b|_{\oalp} \else
{{$\tilde{A}_{FB}^b|_{\oalp}$}}\fi\chkspace}
\def\pafbbso{\relax\ifmmode \tilde{A}_{FB}^b|_{\oalpsq} \else
{{$\tilde{A}_{FB}^b|_{\oalpsq}$}}\fi\chkspace}
\def\pafbbto{\relax\ifmmode \tilde{A}_{FB}^b|_{\oalpc} \else
{{$\tilde{A}_{FB}^b|_{\oalpc}$}}\fi\chkspace}

\def\afbo0{\tilde{A}_{FB}|_{O(0)}}
\def\afbo1{\tilde{A}_{FB}|_{\oalp}}
\def\afbo2{\tilde{A}_{FB}|_{\oalpsq}}
\def\afbo3{\tilde{A}_{FB}|_{\oalpc}}

\def\lam{\relax\ifmmode \Lambda_{\bar{MS}}
       \else {{$\Lambda_{\bar{MS}}$}}\fi\chkspace}
\def\lamuds{\relax\ifmmode \Lambda^{(3)}_{\bar{MS}}
       \else {{$\Lambda^{(3)}_{\bar{MS}}$}}\fi\chkspace}
\def\lamudsc{\relax\ifmmode \Lambda^{(4)}_{\bar{MS}}
       \else $\Lambda^{(4)}_{\bar{MS}}$\fi\chkspace}
\def\lamudscb{\relax\ifmmode \Lambda^{(5)}_{\bar{MS}}
       \else $\Lambda^{(5)}_{\bar{MS}}$\fi\chkspace}

\def\alp{\relax\ifmmode \alpha_s\else $\alpha_s$\fi\chkspace}
\def\alpbar{\relax\ifmmode \bar{\alpha_s}
       \else $\bar{\alpha_s}$\fi\chkspace}
\def\alpmz{\relax\ifmmode \alpha_s(M_Z)\else $\alpha_s(M_Z)$\fi\chkspace}
\def\alpmzsq{\relax\ifmmode \alpha_s(M_Z^2)
       \else $\alpha_s(M_Z^2)$\fi\chkspace}

\def\oalp{\relax\ifmmode O(\alpha_s)\else{{O($\alpha_s$)}}\fi\chkspace}
\def\oalpsq{\relax\ifmmode O(\alpha_s^2)
           \else{{O($\alpha_s^2$)}}\fi\chkspace}
\def\oalpc{\relax\ifmmode O(\alpha_s^3)
           \else{{O($\alpha_s^3$)}}\fi\chkspace}
\def\oalpf{\relax\ifmmode O(\alpha_s^4)
           \else{{O($\alpha_s^4$)}}\fi\chkspace}

\def\rb{\relax\ifmmode R_3^b/R_3^{all}
           \else{{$R_3^b/R_3^{all}$}}\fi\chkspace}
\def\rc{\relax\ifmmode R_3^c/R_3^{all}
           \else{{$R_3^c/R_3^{all}$}}\fi\chkspace}
\def\ruds{\relax\ifmmode R_3^{uds}/R_3^{all}
           \else{{$R_3^{uds}/R_3^{all}$}}\fi\chkspace}
\def\ri{\relax\ifmmode R_3^i/R_3^{all}
           \else{{$R_3^i/R_3^{all}$}}\fi\chkspace}
\def\rj{\relax\ifmmode R_3^j/R_3^{all}
           \else{{$R_3^j/R_3^{all}$}}\fi\chkspace}
\def\alpi{\relax\ifmmode \alpha^i_s/\alpha^{all}_s
           \else{{$\alpha^i_s/\alpha^{all}_s$}}\fi\chkspace}

\def\z0{{$Z^0$}\chkspace}
\def\Dst{\relax\ifmmode {\rm D}^* \else {D$^*$}\fi\chkspace}
\def\Dpl{\relax\ifmmode {\rm D}^+ \else {D$^+$}\fi\chkspace}
\def\D0{\relax\ifmmode {\rm D}^0 \else {D$^0$}\fi\chkspace}
\def\Kst{\relax\ifmmode {\rm K}^* \else {K$^*$}\fi\chkspace}
\def\K0{\relax\ifmmode {\rm K}^0_s \else {K$^0_s$}\fi\chkspace}
\def\Kpl{\relax\ifmmode {\rm K}^+ \else {K$^+$}\fi\chkspace}
\def\Kstz{\relax\ifmmode {\rm K}^{*0} \else {K$^{*0}$}\fi\chkspace}

\def\PRD{{\em Phys. Rev.} D}
\def\ZPC{{\em Z. Phys.} C}

\def\ra{{$\rightarrow$}\chkspace}

\def\be{\begin{equation}}
\def\ee{\end{equation}}
\def\bea{\begin{eqnarray}}
\def\eea{\end{eqnarray}}

\def\z0{$Z^0$}
\def\xb{{$x_B$}\chkspace}

\def\etal{{\it et al.}\chkspace}
\def\adhoc{{\it ad hoc}\chkspace}

\def\qqg{{$q\bar{q}g$}\chkspace}

\def\Zbb{{$Z^0 \rightarrow b\:{\bar b}$}\chkspace}

\renewcommand{\baselinestretch}{1.}

 1

\catcode`\@=11 
\makeatletter
\def\@seccntformat#1{\csname the#1\endcsname.\hskip 1em}

\makeatother
\begin{document}
\thispagestyle{empty}
\begin{flushright}
{\footnotesize\renewcommand{\baselinestretch}{.75}
  SLAC--PUB--8075\\
\vspace{0.1cm}
May 1999\\
}
\end{flushright}

\vskip 0.5truecm
 
\begin{center}
 {\Large \bf An Improved Measurement of\\
the $b$ Quark Fragmentation Function in \z0 Decays$^*$}
 
\vspace {1.4cm}
  {\bf Danning Dong}
\vspace{0.3cm}\\
Massachusetts Institute of Technology, Cambridge, MA 02139\\
\vspace{0.5cm}
{\it Representing} \\
\vspace{0.3cm}
{\bf The SLD Collaboration\\}
\vspace{0.2cm}
Stanford Linear Accelerator Center\\
Stanford University, Stanford, CA 94309
\end{center}
 
\normalsize
 
\vspace{1cm} 
\begin{center}
{\bf Abstract }
\end{center}

{
\linewidth = 100mm
\indent
We present preliminary results of a new measurement 
of the $b$ quark fragmentation function in $Z^{0}$ 
decays using a novel kinematic $B$ hadron energy 
reconstruction technique.  The measurement is performed
using 150,000 hadronic \z0 events recorded in the SLD experiment 
at SLAC between 1996 and 1997.  The small and stable SLC beam spot 
and the CCD-based vertex detector are used to reconstruct 
topological $B$-decay vertices with high efficiency and purity, 
and to provide precise measurements of the kinematic
quantities used in this technique.  We measure the $B$ energy 
with good efficiency and resolution over the full kinematic range.  
We compare the scaled B hadron energy distribution with several 
functional forms of the $B$ hadron energy distribution and predictions 
of several models of $b$ quark fragmentation.  Several functions 
including JETSET $+$ Peterson are excluded by the data.  The average 
scaled energy of the weakly decaying $B$ hadron is measured to be 
\xb $=$ 0.719 $\pm$ 0.005 (stat) $\pm$ 0.007 (syst) 
$\pm$ 0.001 (model) (preliminary).  
}

\vfil
 
\noindent
\begin{center}
{\it Presented at the American Physical Society (APS) \\
Meeting of the Division of Particles and Fields (DPF 99)\\
University of California, Los Angeles\\
January 5-9, 1999}

\end{center}
\vskip .3truecm
\noindent
$^*$ Work supported in part by Department of Energy contracts
DE-FC02-94ER40818 and DE-AC03-76SF00515.
\eject
  
\section{Introduction}

The fragmentation process which transforms colored partons into
colorless hadrons is typically characterized by the fragmentation function.
The $b$ quark fragmentation is of special importance in the study of
quark fragmentation because the large $b$ quark mass provides
a natural mass scale in QCD calculations and allows the application of 
perturbative QCD and heavy quark expansion, and can help to extract 
non-perturbative effects in fragmentation, which is the least understood
part of the fragmentation process.

According to the factorization theorem, the heavy quark 
fragmentation function can be described as a convolution of perturbative
and non-perturbative effects.  For the $b$ quark, 
the perturbative calculation is in principle 
understood~\cite{mn,jaffe,lisa,bcfy,dkt}.
Nonperturbative effects have been parametrized in 
both model-dependent~\cite{kart,bowler,pete,lund,collins,mn} and 
model-independent approaches~\cite{jaffe,lisa,bcfy,dkt}.
The fact that several models are yet to be 
experimentally tested is an indication of a lack of 
precise and conclusive experimental results, if not theoretical 
understanding.

It is indeed experimentally challenging to measure 
the $b$ quark fragmentation function to a level of 
precision sufficient to distinguish among the various models.
Since the $b$ quark fragmentation function is the probability 
distribution of the fraction of the momentum of the $b$ quark 
carried by the $B$ hadron, the most sensitive experimental determination 
of the shape of the $b$ fragmentation function is expected to come 
from a precise direct measurement of the 
$B$ hadron energy (or momentum) distribution.  
The difficulty in precisely measuring the $B$ hadron energy
distribution stems mostly from the fact that 
most of the $B$ decays can only be partially reconstructed,
causing a significant fraction of the $B$ energy to be missing 
from the $B$ decay vertex.
Recent direct measurements at LEP~\cite{delphi93,aleph95} 
and SLD~\cite{sld96} have used overall energy-momentum constraints and 
calorimetric information to extract this missing energy in a 
sample of semi-leptonic $B$ decays.  These 
measurements suffer from low 
statistics as well as poor $B$ energy resolution at low energy,
and hence have a relatively weak discriminating power between different shapes 
of the fragmentation function.
Indirect measurements~\cite{early} such as the measurements of the 
lepton spectrum and charged multiplicity have been used to constrain the average $B$ energy.  However, these measurements are 
not very sensitive to the shape of the energy distribution.  

Here we report preliminary results of SLD's new measurement of the
$B$ hadron energy distribution.  
We developed a novel kinematic technique 
which uses only charged tracks associated with the 
$B$ vertex and the $B$ flight direction to reconstruct 
individual $B$ hadron energy with good resolution over 
the full kinematic range while achieving an efficiency
much higher than previous measurements.

\section{$B$ hadron Selection}
A general description of the SLD detector can be found 
elsewhere~\cite{sld1,sld2}
The excellent tracking and vertexing capabilities at SLD \cite{vxd3}
are exploited in the reconstruction of $B$ decays in \Zbb events.

A set of cuts is applied to select hadronic \z0 events well-contained 
within the detector acceptance.
The efficiency for selecting a well-contained \z0 \ra $q\bar{q}(g)$
event is estimated to be above 96\% independent of quark flavor. The
selected sample comprise 111,569 events, with an estimated
$0.10 \pm 0.05\%$ background contribution dominated
by $Z^0 \rightarrow \tau^+\tau^-$ events.

The $B$ sample is selected using a topological vertexing
technique based on the detection and measurement of charged tracks,
which is described in full detail in Ref.~\cite{zvnim}.  
The topological vertexing algorithm~\cite{zvnim} is applied separately to 
the set of ``quality'' tracks in each hemisphere (defined with 
respect to the event thrust axis).  

When a candidate vertex is found, tracks not associated 
with this ``seed'' vertex are attached to the vertex if they are 
more likely to have originated from this vertex than from the IP.  
This track-attachment procedure is tuned to minimize false 
track-vertex associations to the vertex.  
Attaching a false track to the 
vertex affects the vertex-kinematics more than 
failing to associate a genuine track originated from the vertex, 
and hence can cause significant degradation in the reconstructed 
$B$ energy resolution.  
On average, this procedure attaches 0.8 tracks to each seed vertex;
about 92\% of the reconstructed tracks which originated from the $B$-decay
are associated with the reconstructed vertex, and 98\% of the 
vertex-associated tracks are true $B$ decays tracks.

\parbox[b]{2.7 in}{
\epsfxsize 2.4 in
\epsfbox{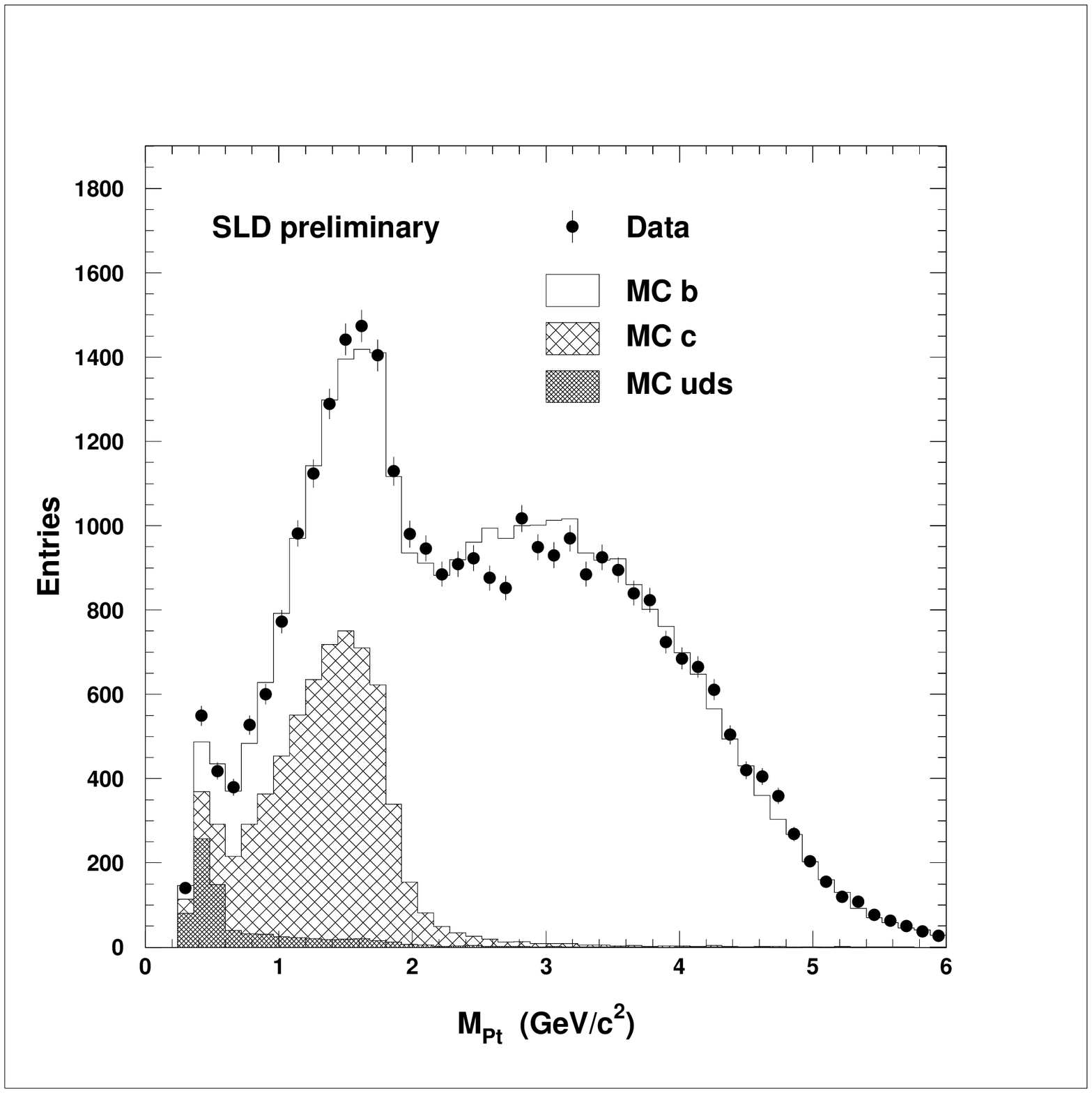}
\vspace{0.3cm}
Figure 1 Distribution of the reconstructed $P_{t}$-corrected vertex mass in
 the 1996-97 data (points).  Also shown is the prediction of the Monte
 Carlo simulation, for which the flavor composition is indicated.
\vspace{0.3cm}
}
\ \hspace{0.1cm} \ 
\parbox[b]{3.15in} { \vspace{0.1cm}
\hspace{1.3em} 
The mass of the reconstructed vertex, $M_{ch}$, is calculated by 
assigning each track the charged-pion mass.  
Because of the tiny SLC IP error and the excellent 
vertex resolution, the $B$ flight direction pointing
along the line joining the IP and the secondary vertex 
is well-measured.  
Therefore the transverse momentum $P_t$ of tracks associated with 
the vertex relative to the $B$ flight direction is also well-measured.  
The mass of the missing particles can then 
be partially compensated by using $P_t$ to form the 
``$P_t$ corrected mass'', $M_{P_t} = \sqrt{M_{ch}^2 + P_t^2} + |P_t|$. 
To minimize effects of large fluctuations of $P_t$ at short decay
length, 
the minimum transverse momentum (which is varied within the $1\sigma$ limits
 constraining the axis at the measured 
 interaction point (IP) and reconstructed seed vertex)
 is calculated in order to determine $M_{P_t}$.
Figure 1 shows the distribution of the $P_t$-corrected 
mass (points) for the 32,492 accepted hemi-
}
spheres in the data sample, and
 the corresponding simulated distribution.  
To obtain a high purity $B$ sample, 
$B$ hadron candidates are selected by requiring 
$M_{P_t}$ $>$ 2.0 GeV/$c^{2}$.  
A total of 19,604 hemispheres are selected, 
with an estimated efficiency for selecting a true $B$-hemisphere 
of 40.1\%, and a sample purity of 98.2\%.  The contributions from 
light-flavor events in the sample are 0.15\% for primary u,d and s events
and 1.6\% for c events.

\section{$B$ Energy Reconstruction}
\vspace{-0.1cm}
Since the sum of the charged track energy at the $B$ vertex, $E_{ch}$, is
known, we are only concerned with finding the energy of particles missing 
from the $B$ vertex.

\parbox[b]{2.85in}{
\leavevmode
\epsfxsize 2.85in
\epsfbox{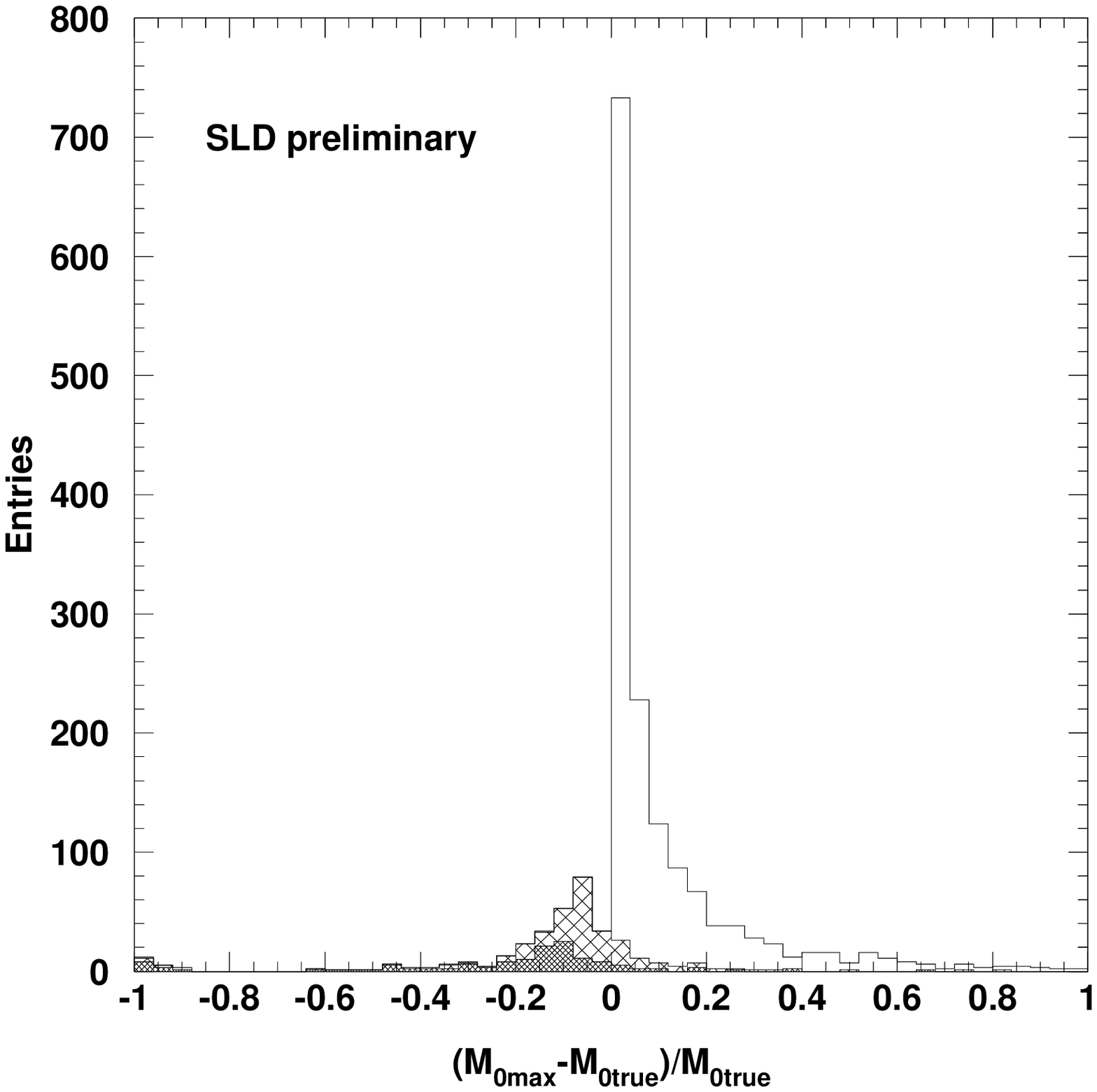}
\small 
\vspace{0.1cm}
Figure 2. The relative deviation of the maximum missing mass from the true 
missing mass for Monte Carlo simulated $B$ hadron decays, which is divided
into three categories: $B^0$ and $B^{\pm}$ (open), 
$B_s^0$ (cross-hatched), and $\Lambda_b$ (dark hatched). 
\vspace{0.2cm}
} \
\hspace{0.2cm} \
\parbox[b]{2.9in}{\vspace{-0.1cm}
\leavevmode
\epsfxsize 2.9in
\epsfxsize 2.85in
\epsfbox{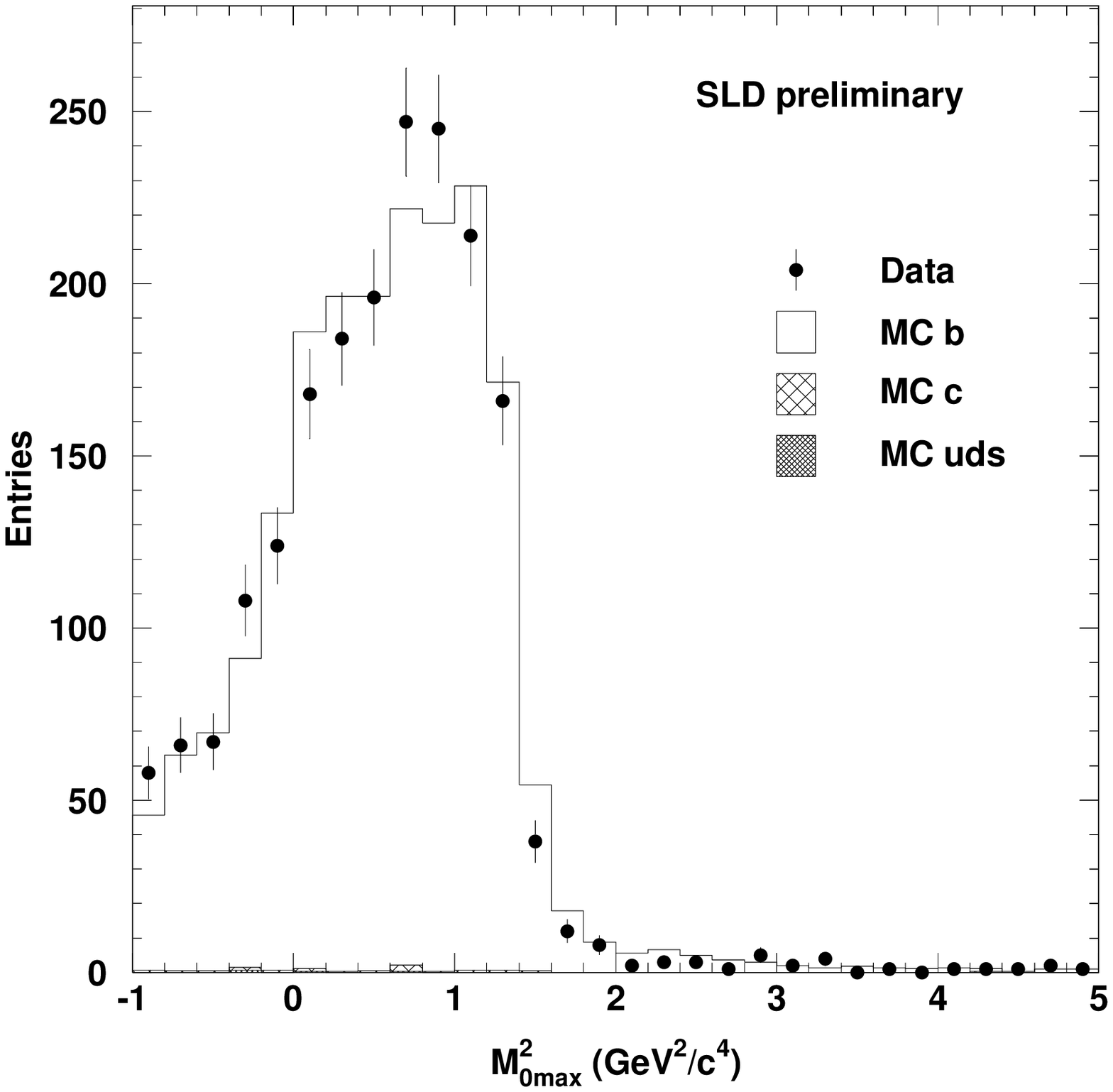}
\small 
\vspace{0.1cm}
Figure 3.
Distribution of the reconstructed $M_{0max}^{2}$ for the selected 
vertices in the 1996-97 data (points).  
Also shown is the prediction of the Monte Carlo simulation.
\vspace{1.3cm}
}
\indent
Given reconstructed $B$ vertex, an upper bound 
on the mass of the missing particles from the vertex is found to be
$M_{0max}^2 = M_B^2 - 2M_B\sqrt{M_{ch}^2 + P_t^2} + M_{ch}^2, $
where we assume the true mass of the $B$ hadron decayed
at the vertex, $M_B$, equals the $B^0$ meson mass.  
Since the true missing mass $M_0^{true}$ is 
often rather close to $M_{0max}$ (Figure 2),
$M_{0max}$ is subsequently used as an estimate of $M_0^{true}$ 
($M_0^{true}$ is set to 0 if the reconstructed $M_{0max}$ is negative)
to solve
 for the longitudinal momentum of the missing particles from kinematics:
\begin{center}
$P_{0l} = \frac {\textstyle M_{B}^{2}-(M_{ch}^{2}+P_{t}^{2})-(M_{0}^{2}+P_{t}^{2})}{\textstyle 2 (M_{ch}^{2}+P_{t}^{2})} P_{chl}$,
\end{center}
and hence the missing $B$ energy from the vertex, $E_0$.   
The $B$ hadron energy is then $E_B = E_{ch} + E_0$.
Since $ 0 \leq  M_0^{true} \leq M_{0max}$, 
the $B$ energy is well-constrained when $M_{0max}$ is small.  In addition,
most $uds$ and $c$ backgrounds are concentrated at large
$M_{0max}$.  
We choose an {\it ad hoc} upper cut on the $M_{0max}^2$
to achieve a nearly $x_B$-independent $B$ selection
 efficiency.  
Figure 3 shows the distribution of $M_{0max}^{2}$ after 
these cuts, where the data and Monte Carlo simulation are in good agreement.
A total of 1920 vertices in the 1996-97 data satisfy all selection cuts.
Figure 4 shows the distribution of the reconstructed scaled weakly
decaying $B$ hadron energy for data and Monte Carlo.
The overall $B$ selection efficiency is 3.9\% and 
the estimated purity is about 99.5\%.  The efficiency as a function of 
$x_B^{true}$ is shown in Figure 5.  
We examine the normalized difference between the true and reconstructed 
$B$ hadron energies for Monte Carlo events.  The distribution is fitted 
by a double Gaussian, resulting in a core width 
(the width of the narrower Gaussian) of 10.4\% and a tail width 
(the width of the wider Gaussian) of 23.6\% with a core fraction of 83\%.  
Figure 6 shows the core and tail widths as a function 
of $x_{B}^{true}$.  The core width depends only weakly on 
the true $x_B$, another feature that makes this method unique.  
\\
\parbox[b]{6.0in}{ 
\epsfxsize 6.0 in
\epsfysize 5.0 in
\epsffile{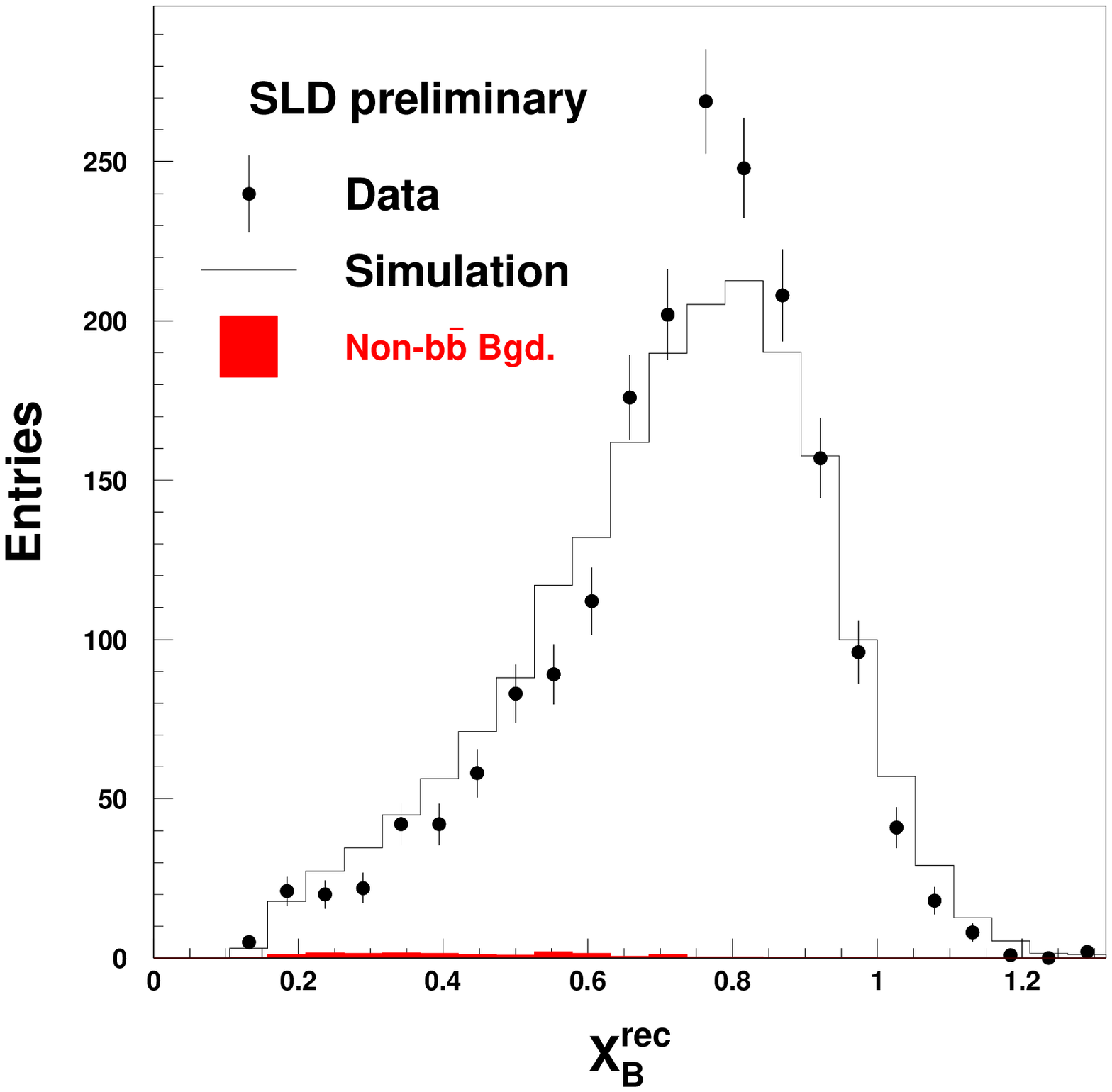}
{  \vspace{-0.3cm}\small Figure 4.
Distribution of the reconstructed scaled $B$ hadron energy for data(points)
and simulation (histogram).  The solid histogram shows the 
non-$b\bar{b}$ background.
  \vspace{0.2cm}
}
}

\parbox[b]{2.9 in}{
  \vspace{-0.1cm}
  \epsfxsize 2.8in
  \epsffile{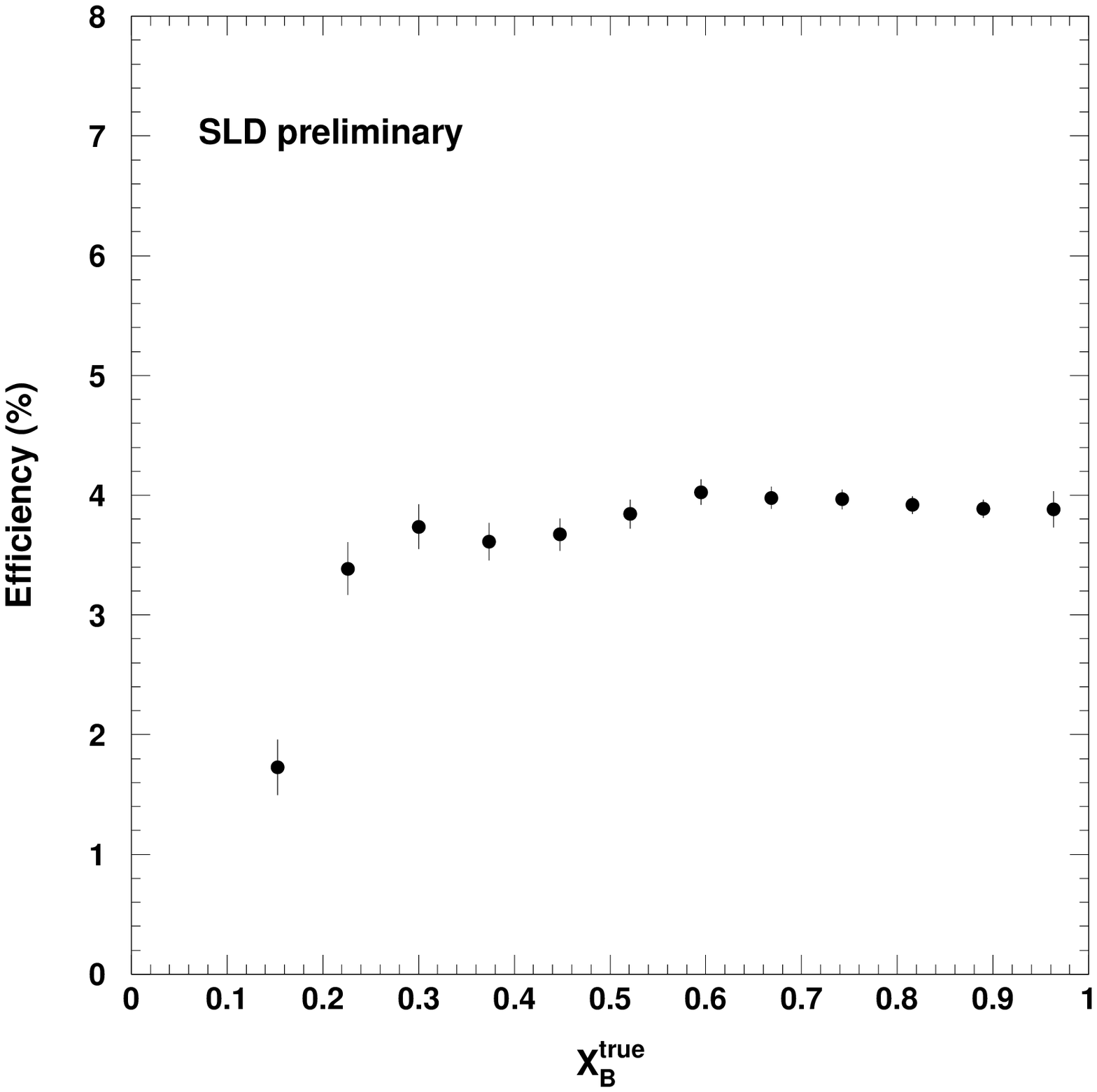}
{\small Figure 5.
Distribution of the efficiency as a function of the true $B$ energy.
\vspace{0.4cm}}
} 
\ \hspace{0.05cm} \
\parbox[b]{2.9in}{ 
 \epsfxsize 2.8 in
  \epsffile{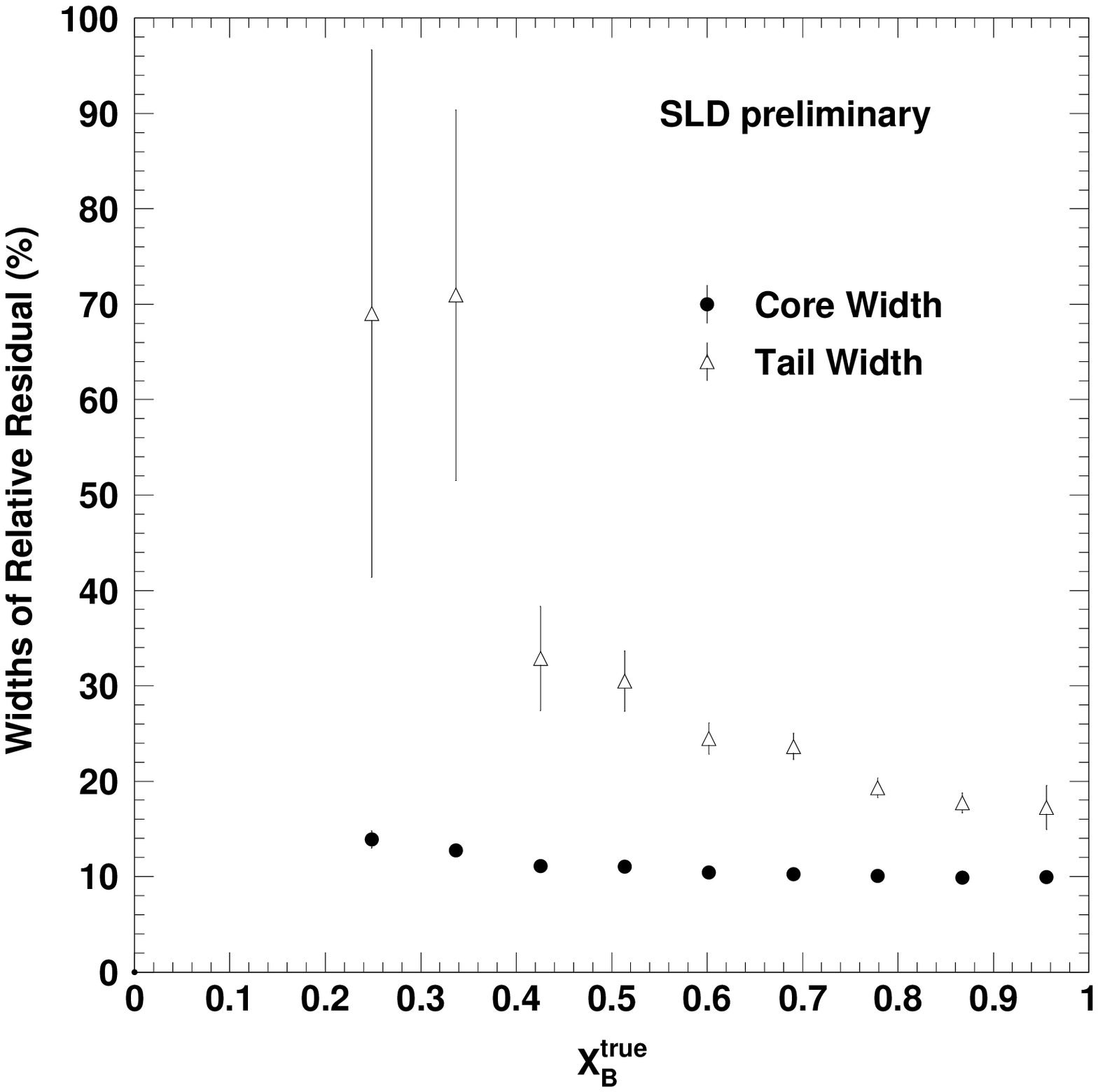}
{\small Figure 6.
The fitted core and tail widths of the $B$ energy resolution as
 a function of the true scaled $B$ hadron energy.
}
}
\section{Tests of Functional Forms and Models}
After background subtraction, 
the distribution of the reconstructed scaled $B$ hadron energy  
is compared with a set of {\em ad hoc} functional forms of the 
$x_B$ distribution in order to estimate the variation 
in the shape of the $x_B$ distribution. 
For each functional form, the default SLD Monte Carlo is re-weighted 
and then compared with the data bin-by-bin and a $\chi^2$ is computed.  
The minimum $\chi^2$ is found by varying the input parameter(s).
The Peterson function~\cite{pete}, two
 \adhoc generalizations of the 
Peterson function\cite{aleph95} (ALEPH 1 and 2)
and a 7th-order polynomial~\footnote{The behavior of 
this polynomial is rather unphysical at low $x_B$ and will not be considered 
hereafter} are consistent with the data.  
We exclude the functional forms described in 
BCFY~\cite{bcfy}, Collins and Spiller~\cite{collins}, 
Kartvelishvili~\cite{kart}, Lund~\cite{lund} and 
a power function of the form $f(x)=x^\alpha(1-x)^\beta$.
The result is shown in Figure 6 and in Table 1 and 2.

\parbox[b]{5.9 in}{
 \vspace{0.1cm}
  \epsfxsize 5.9 in
  \epsfysize 5.4 in
  \epsffile{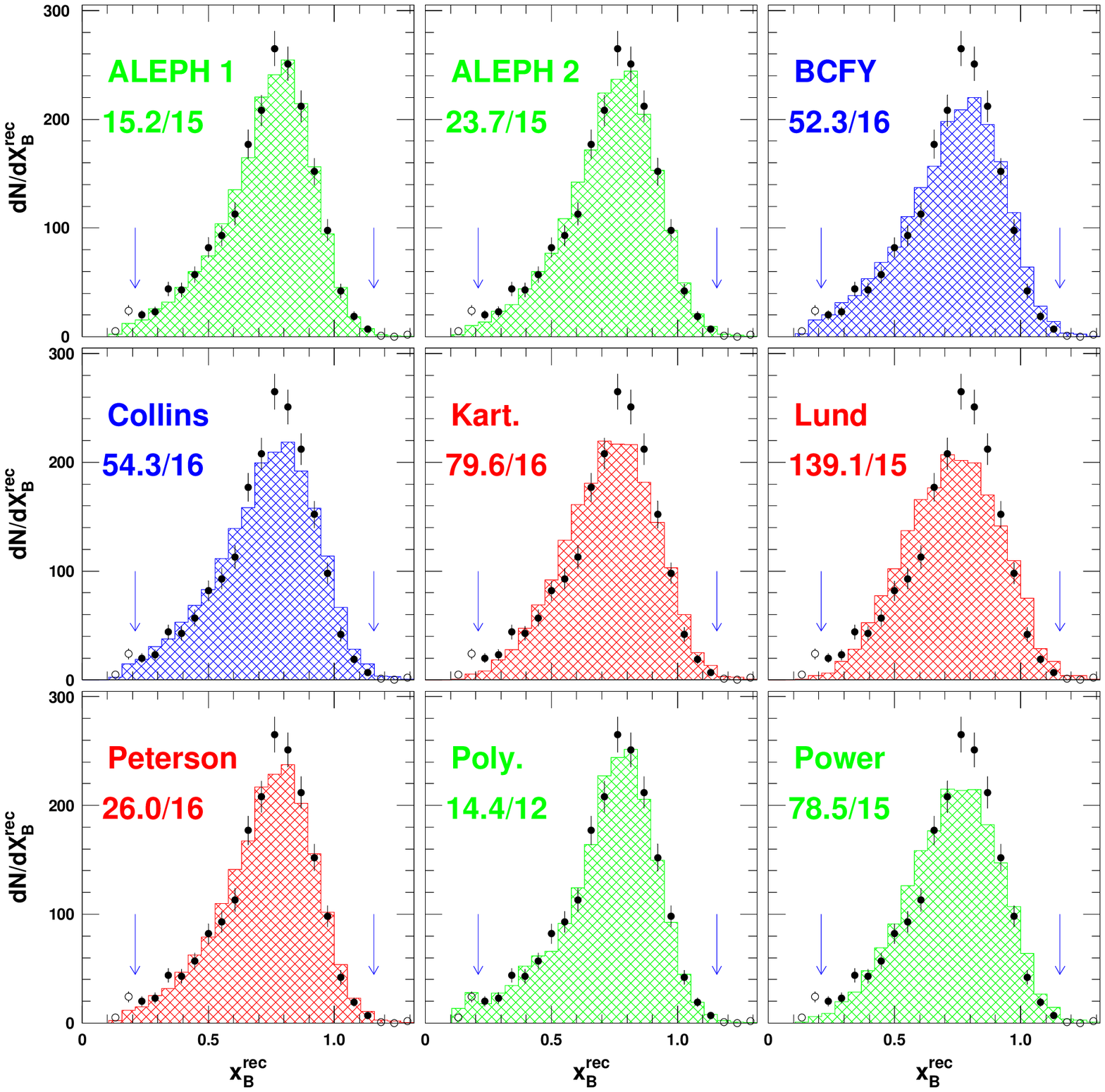}
{\small Figure 7.
Each figure shows the background-subtracted distribution of 
reconstructed $B$ hadron energy for the data (points) and for the simulation 
(histograms) based on the respective optimised input fragmentation function.  
The $\chi^2$ fit uses data in the bins between the 
two arrows. \vspace{0.1cm}}}
\\
\indent


\begin{tabular}{lcccc} \\
\hline
Function  &  $D(x)$    &   Reference \\
\hline
\vspace{0.1cm}
ALEPH 1 & $\frac{\textstyle 1-x}{\textstyle x}(1-\frac{\textstyle c}{\textstyle x}-\frac{\textstyle d}{\textstyle 1-x})^{-2}$ & \cite{aleph95} \\
ALEPH 2 & $\frac{\textstyle 1}{\textstyle x}(1-\frac{\textstyle c}{\textstyle x}-\frac{\textstyle d}{\textstyle 1-x})^{-2}$ & \cite{aleph95} \\
\vspace{1 mm}
BCFY &  $\frac{\textstyle x(1-x)^{2}}{\textstyle [1-(1-r)x]^{6}}[3+{\sum_{i=1}^{4} (-x)^{i}f_{i}(r)}]$  & \cite{bcfy} \\
\vspace{1 mm}
Collins and Spiller &$ (\frac{\textstyle 1-x}{\textstyle x}+\frac{\textstyle (2-x)\epsilon_{b}}{\textstyle 1-x})
(1+x^{2})(1-\frac{\textstyle 1}{\textstyle x}-\frac{\textstyle \epsilon_{b}}{\textstyle 1-x})^{-2}$ & \cite{collins} \\
\vspace{1 mm}
Kartvelishvili {\em et al.}  & $x^{\alpha_{b}}(1-x)$ & \cite{kart} \\
\vspace{1 mm}
Lund  & $\frac{\textstyle 1}{\textstyle x}(1-x)^{a}exp(-bm_{\perp}^{2}/x)$
      & \cite{lund} \\
\vspace{1 mm}
Peterson {\em et al.}  & $\frac{\textstyle 1}{\textstyle x}(1-\frac{\textstyle 1}{\textstyle x}-\frac{\textstyle \epsilon_{b}}{\textstyle 1-\textstyle x})^{-2}$   & \cite{pete} \\
\vspace{1 mm}
Polynomial & $x(1-x)(1+{\sum_{i=1}^{5} p_{i}x^{i}})$  &   (see text)  \\
\vspace{1 mm}
Power  & $x^{\alpha}(1-x)^{\beta}$  &  (see text) \\
\hline
\vspace{0.2cm}
\end{tabular}
\\
\noindent 
Table 1. Fragmentation functional forms used in comparison with the data.  
For the BCFY function $f_{1}(r)~=~3(3-4r)$, 
$f_{2}(r)~=~12-23r+26r^{2}$, $f_{3}(r)~=~(1-r)(9-11r+12r^{2})$, and
$f_{4}(r)~=~3(1-r)^{2}(1-r+r^{2})$.  A polynomial function and 
a power function are also included.  


\vspace{0.3cm}
\begin{center}
\begin{tabular}{lcccc}
\hline
Function  &  $\chi^{2}/dof$    &   Parameters &  $\langle x_{B} \rangle$\\
\hline
ALEPH 1 & 15.2/15  & $c=0.860_{-0.018}^{+0.019}$ &    0.718$\pm$0.005 \\     
      &            & $d=0.019\pm0.002$           &     \\
\vspace{1 mm}
ALEPH 2 & 23.7/15  & $c=0.938_{-0.034}^{+0.039}$ & 0.720$\pm$0.005 \\ 
      &            & $d=0.036\pm0.002$           &     \\
\vspace{1 mm}
BCFY  &  52.3/16   & $r=0.2316_{-0.0088}^{+0.0092}$ & 0.713$\pm$0.005 \\
\vspace{1 mm}
Collins and Spiller  
     &  54.3/16   & $\epsilon_{b}=0.044_{-0.004}^{+0.005}$ & 0.714$\pm$0.005 \\
\vspace{1 mm}
Kartvelishvili {\em et al.}  
      &  79.6/16   & $\alpha_{b}=4.15\pm0.11$  &  0.720$\pm$0.004   \\
\vspace{1 mm}
Lund  &  139.1/15  & $a=2.116_{-0.114}^{+0.118}$ & 0.720$\pm$0.005 \\
      &            & $bm_{\perp}^{2}=0.408_{-0.070}^{+0.073}$ &     \\
\vspace{1 mm}
Peterson {\em et al.}  & 26.0/16 & 
$\epsilon_{b}=0.0338_{-0.0022}^{+0.0020}$ & 0.719$\pm$0.005 \\
\vspace{1 mm}
Polynomial
      & 14.4/12    &  $p_{1}=-12.4\pm0.4$             &  (see text)       \\  
      &            &  $p_{2}=58.7\pm1.9$    &   \\
            &          & $p_{3}=-130.5\pm4.2$  &   \\
            &          & $p_{4}=136.8\pm4.3$   &   \\
            &          & $p_{5}=-53.7\pm1.8$   &   \\
\vspace{1 mm}
Power 
      &  78.5/15   & $\alpha=3.91_{-0.24}^{+0.25}$   & 0.722$\pm$0.005  \\
      &            & $\beta=0.894_{-0.097}^{+0.102}$  &                  \\
\hline
\end{tabular}
\end{center}

\vspace{0.3cm}
\noindent 
Table 2. Results of the $\chi^{2}$ fit of fragmentation functions 
to the reconstructed
$B$ hadron energy distribution after background subtraction.  Minimum 
$\chi^{2}$, number of degrees of freedom and coresponding parameter
values are listed.  Errors are statistical only. 
\vspace{0.3cm}

\indent
We then test several heavy quark fragmentation models.
Since the fragmentation functions 
are usually functions of an 
experimentally inaccessible variable (e.g. $z=(E+p_{\|})_{H}/(E+p_{\|})_Q$), 
it is necessary to use a Monte Carlo generator 
such as JETSET~\cite{jetset} 
to generate events according to a given input heavy
quark fragmentation function.  The resulting $B$ energy distribution is
then used to re-weigh the Monte Carlo distribution before 
comparing with the data. 
The minimum $\chi^2$ is found by varying the input parameter(s).
Within the context of the JETSET Monte Carlo, 
Bowler~\cite{bowler} and the Lund~\cite{lund}
models are consistent with the data, while 
Peterson~\cite{pete} model is found to be inconsistent with the data.

\section{Systematic Errors}
We have considered both detector and physics modelling systematics. 
The dominant systematic error is  related to 
charged track transverse momentum resolution smearing, which has been
evaluated conservatively and can be reduced with a detailed study. 
All physics systematics are rather small.  
Other relevant systematic effects such as by varying 
the event selection cuts and the assumed $B$ hadron mass are also 
found to be small.  In each case, 
conclusions about the shape of the $B$ energy distribution hold, and 
the systematics in the average $B$ hadron energy is added in quadrature
to obtain the total systematics.

\section{Conclusion}

Taking advantage of SLC's small beam-spot and SLD's high vertex resolution,
we have developed a new kinematic technique 
to measure, for the first time, the $B$ hadron energy distribution in
\z0 decays with good resolution over the full kinematic range.
Using 1996-97 data, 
we exclude several functional forms of the $B$ energy distribution 
and the JETSET + Peterson fragmentation model.  
The mean of the scaled weakly decaying $B$ hadron energy distribution 
is obtained by taking the average of the means of the three functional
forms which are found to be consistent with the data.  The r.m.s.\ of 
the three means is regarded as a minimum error on model-dependence.  We find
\begin{center}
$<x_B> = 0.719\pm 0.005 (stat)\pm0.007 (syst)\pm0.001 (model)$
\end{center}
\noindent where the small model-dependence error
indicates that $<x_B>$ is relatively insensitive 
to the allowed forms of the shape of the fragmentation function. 
 The precision in 
the measured average $B$ hadron energy represents a substantial improvement
over previous direct measurements.  All results are preliminary.

\vskip 1truecm
  
\section*{$^{**}$List of Authors} 
%
%
%
\begin{center}
\def\iADEL{$^{(1)}$}
\def\iAOMORI{$^{(2)}$}
\def\iBOLO{$^{(3)}$}
\def\iBRUN{$^{(4)}$}
\def\iBU{$^{(5)}$}
\def\iCINC{$^{(6)}$}
\def\iCOLO{$^{(7)}$}
\def\iCOLU{$^{(8)}$}
\def\iCSU{$^{(9)}$}
\def\iFERR{$^{(10)}$}
\def\iFRAS{$^{(11)}$}
\def\iILLI{$^{(12)}$}
\def\iLBL{$^{(13)}$}
\def\iLTU{$^{(14)}$}
\def\iMASS{$^{(15)}$}
\def\iMISSI{$^{(16)}$}
\def\iMIT{$^{(17)}$}
\def\iMOSCOW{$^{(18)}$}
\def\iNAGO{$^{(19)}$}
\def\iOREG{$^{(20)}$}
\def\iOXF{$^{(21)}$}
\def\iPADO{$^{(22)}$}
\def\iPERU{$^{(23)}$}
\def\iPISA{$^{(24)}$}
\def\iRAL{$^{(25)}$}
\def\iRUTG{$^{(26)}$}
\def\iSLAC{$^{(27)}$}
\def\iSOGA{$^{(28)}$}
\def\iSOONG{$^{(29)}$}
\def\iTENN{$^{(30)}$}
\def\iTOHO{$^{(31)}$}
\def\iUCSB{$^{(32)}$}
\def\iUCSC{$^{(33)}$}
\def\iVAND{$^{(34)}$}
\def\iWASH{$^{(35)}$}
\def\iWISC{$^{(36)}$}
\def\iYALE{$^{(37)}$}

  \baselineskip=.75\baselineskip  
\mbox{K. Abe\unskip,\iAOMORI}
\mbox{K.  Abe\unskip,\iNAGO}
\mbox{T. Abe\unskip,\iSLAC}
\mbox{I.Adam\unskip,\iSLAC}
\mbox{T.  Akagi\unskip,\iSLAC}
\mbox{N. J. Allen\unskip,\iBRUN}
\mbox{A. Arodzero\unskip,\iOREG}
\mbox{W.W. Ash\unskip,\iSLAC}
\mbox{D. Aston\unskip,\iSLAC}
\mbox{K.G. Baird\unskip,\iMASS}
\mbox{C. Baltay\unskip,\iYALE}
\mbox{H.R. Band\unskip,\iWISC}
\mbox{M.B. Barakat\unskip,\iLTU}
\mbox{O. Bardon\unskip,\iMIT}
\mbox{T.L. Barklow\unskip,\iSLAC}
\mbox{J.M. Bauer\unskip,\iMISSI}
\mbox{G. Bellodi\unskip,\iOXF}
\mbox{R. Ben-David\unskip,\iYALE}
\mbox{A.C. Benvenuti\unskip,\iBOLO}
\mbox{G.M. Bilei\unskip,\iPERU}
\mbox{D. Bisello\unskip,\iPADO}
\mbox{G. Blaylock\unskip,\iMASS}
\mbox{J.R. Bogart\unskip,\iSLAC}
\mbox{B. Bolen\unskip,\iMISSI}
\mbox{G.R. Bower\unskip,\iSLAC}
\mbox{J. E. Brau\unskip,\iOREG}
\mbox{M. Breidenbach\unskip,\iSLAC}
\mbox{W.M. Bugg\unskip,\iTENN}
\mbox{D. Burke\unskip,\iSLAC}
\mbox{T.H. Burnett\unskip,\iWASH}
\mbox{P.N. Burrows\unskip,\iOXF}
\mbox{A. Calcaterra\unskip,\iFRAS}
\mbox{D.O. Caldwell\unskip,\iUCSB}
\mbox{D. Calloway\unskip,\iSLAC}
\mbox{B. Camanzi\unskip,\iFERR}
\mbox{M. Carpinelli\unskip,\iPISA}
\mbox{R. Cassell\unskip,\iSLAC}
\mbox{R. Castaldi\unskip,\iPISA}
\mbox{A. Castro\unskip,\iPADO}
\mbox{M. Cavalli-Sforza\unskip,\iUCSC}
\mbox{A. Chou\unskip,\iSLAC}
\mbox{E. Church\unskip,\iWASH}
\mbox{H.O. Cohn\unskip,\iTENN}
\mbox{J.A. Coller\unskip,\iBU}
\mbox{M.R. Convery\unskip,\iSLAC}
\mbox{V. Cook\unskip,\iWASH}
\mbox{R. Cotton\unskip,\iBRUN}
\mbox{R.F. Cowan\unskip,\iMIT}
\mbox{D.G. Coyne\unskip,\iUCSC}
\mbox{G. Crawford\unskip,\iSLAC}
\mbox{C.J.S. Damerell\unskip,\iRAL}
\mbox{M. N. Danielson\unskip,\iCOLO}
\mbox{M. Daoudi\unskip,\iSLAC}
\mbox{N. de Groot\unskip,\iSLAC}
\mbox{R. Dell'Orso\unskip,\iPERU}
\mbox{P.J. Dervan\unskip,\iBRUN}
\mbox{R. de Sangro\unskip,\iFRAS}
\mbox{M. Dima\unskip,\iCSU}
\mbox{A. D'Oliveira\unskip,\iCINC}
\mbox{D.N. Dong\unskip,\iMIT}
\mbox{P.Y.C. Du\unskip,\iTENN}
\mbox{R. Dubois\unskip,\iSLAC}
\mbox{B.I. Eisenstein\unskip,\iILLI}
\mbox{V. Eschenburg\unskip,\iMISSI}
\mbox{E. Etzion\unskip,\iWISC}
\mbox{S. Fahey\unskip,\iCOLO}
\mbox{D. Falciai\unskip,\iFRAS}
\mbox{C. Fan\unskip,\iCOLO}
\mbox{J.P. Fernandez\unskip,\iUCSC}
\mbox{M.J. Fero\unskip,\iMIT}
\mbox{K.Flood\unskip,\iMASS}
\mbox{R. Frey\unskip,\iOREG}
\mbox{T. Gillman\unskip,\iRAL}
\mbox{G. Gladding\unskip,\iILLI}
\mbox{S. Gonzalez\unskip,\iMIT}
\mbox{E.L. Hart\unskip,\iTENN}
\mbox{J.L. Harton\unskip,\iCSU}
\mbox{A. Hasan\unskip,\iBRUN}
\mbox{K. Hasuko\unskip,\iTOHO}
\mbox{S. J. Hedges\unskip,\iBU}
\mbox{S.S. Hertzbach\unskip,\iMASS}
\mbox{M.D. Hildreth\unskip,\iSLAC}
\mbox{J. Huber\unskip,\iOREG}
\mbox{M.E. Huffer\unskip,\iSLAC}
\mbox{E.W. Hughes\unskip,\iSLAC}
\mbox{X.Huynh\unskip,\iSLAC}
\mbox{H. Hwang\unskip,\iOREG}
\mbox{M. Iwasaki\unskip,\iOREG}
\mbox{D. J. Jackson\unskip,\iRAL}
\mbox{P. Jacques\unskip,\iRUTG}
\mbox{J.A. Jaros\unskip,\iSLAC}
\mbox{Z.Y. Jiang\unskip,\iSLAC}
\mbox{A.S. Johnson\unskip,\iSLAC}
\mbox{J.R. Johnson\unskip,\iWISC}
\mbox{R.A. Johnson\unskip,\iCINC}
\mbox{T. Junk\unskip,\iSLAC}
\mbox{R. Kajikawa\unskip,\iNAGO}
\mbox{M. Kalelkar\unskip,\iRUTG}
\mbox{Y. Kamyshkov\unskip,\iTENN}
\mbox{H.J. Kang\unskip,\iRUTG}
\mbox{I. Karliner\unskip,\iILLI}
\mbox{H. Kawahara\unskip,\iSLAC}
\mbox{Y. D. Kim\unskip,\iSOGA}
\mbox{R. King\unskip,\iSLAC}
\mbox{M.E. King\unskip,\iSLAC}
\mbox{R.R. Kofler\unskip,\iMASS}
\mbox{N.M. Krishna\unskip,\iCOLO}
\mbox{R.S. Kroeger\unskip,\iMISSI}
\mbox{M. Langston\unskip,\iOREG}
\mbox{A. Lath\unskip,\iMIT}
\mbox{D.W.G. Leith\unskip,\iSLAC}
\mbox{V. Lia\unskip,\iMIT}
\mbox{C.-J. S. Lin\unskip,\iSLAC}
\mbox{X. Liu\unskip,\iUCSC}
\mbox{M.X. Liu\unskip,\iYALE}
\mbox{M. Loreti\unskip,\iPADO}
\mbox{A. Lu\unskip,\iUCSB}
\mbox{H.L. Lynch\unskip,\iSLAC}
\mbox{J. Ma\unskip,\iWASH}
\mbox{G. Mancinelli\unskip,\iRUTG}
\mbox{S. Manly\unskip,\iYALE}
\mbox{G. Mantovani\unskip,\iPERU}
\mbox{T.W. Markiewicz\unskip,\iSLAC}
\mbox{T. Maruyama\unskip,\iSLAC}
\mbox{H. Masuda\unskip,\iSLAC}
\mbox{E. Mazzucato\unskip,\iFERR}
\mbox{A.K. McKemey\unskip,\iBRUN}
\mbox{B.T. Meadows\unskip,\iCINC}
\mbox{G. Menegatti\unskip,\iFERR}
\mbox{R. Messner\unskip,\iSLAC}
\mbox{P.M. Mockett\unskip,\iWASH}
\mbox{K.C. Moffeit\unskip,\iSLAC}
\mbox{T.B. Moore\unskip,\iYALE}
\mbox{M.Morii\unskip,\iSLAC}
\mbox{D. Muller\unskip,\iSLAC}
\mbox{V.Murzin\unskip,\iMOSCOW}
\mbox{T. Nagamine\unskip,\iTOHO}
\mbox{S. Narita\unskip,\iTOHO}
\mbox{U. Nauenberg\unskip,\iCOLO}
\mbox{H. Neal\unskip,\iSLAC}
\mbox{M. Nussbaum\unskip,\iCINC}
\mbox{N.Oishi\unskip,\iNAGO}
\mbox{D. Onoprienko\unskip,\iTENN}
\mbox{L.S. Osborne\unskip,\iMIT}
\mbox{R.S. Panvini\unskip,\iVAND}
\mbox{H. Park\unskip,\iOREG}
\mbox{C. H. Park\unskip,\iSOONG}
\mbox{T.J. Pavel\unskip,\iSLAC}
\mbox{I. Peruzzi\unskip,\iFRAS}
\mbox{M. Piccolo\unskip,\iFRAS}
\mbox{L. Piemontese\unskip,\iFERR}
\mbox{E. Pieroni\unskip,\iPISA}
\mbox{K.T. Pitts\unskip,\iOREG}
\mbox{R.J. Plano\unskip,\iRUTG}
\mbox{R. Prepost\unskip,\iWISC}
\mbox{C.Y. Prescott\unskip,\iSLAC}
\mbox{G.D. Punkar\unskip,\iSLAC}
\mbox{J. Quigley\unskip,\iMIT}
\mbox{B.N. Ratcliff\unskip,\iSLAC}
\mbox{T.W. Reeves\unskip,\iVAND}
\mbox{J. Reidy\unskip,\iMISSI}
\mbox{P.L. Reinertsen\unskip,\iUCSC}
\mbox{P.E. Rensing\unskip,\iSLAC}
\mbox{L.S. Rochester\unskip,\iSLAC}
\mbox{P.C. Rowson\unskip,\iCOLU}
\mbox{J.J. Russell\unskip,\iSLAC}
\mbox{O.H. Saxton\unskip,\iSLAC}
\mbox{T. Schalk\unskip,\iUCSC}
\mbox{R.H. Schindler\unskip,\iSLAC}
\mbox{B.A. Schumm\unskip,\iUCSC}
\mbox{J. Schwiening\unskip,\iSLAC}
\mbox{S. Sen\unskip,\iYALE}
\mbox{V.V. Serbo\unskip,\iWISC}
\mbox{M.H. Shaevitz\unskip,\iCOLU}
\mbox{J.T. Shank\unskip,\iBU}
\mbox{G. Shapiro\unskip,\iLBL}
\mbox{D.J. Sherden\unskip,\iSLAC}
\mbox{K. D. Shmakov\unskip,\iTENN}
\mbox{C. Simopoulos\unskip,\iSLAC}
\mbox{N.B. Sinev\unskip,\iOREG}
\mbox{S.R. Smith\unskip,\iSLAC}
\mbox{M. B. Smy\unskip,\iCSU}
\mbox{J.A. Snyder\unskip,\iYALE}
\mbox{H. Staengle\unskip,\iCSU}
\mbox{A. Stahl\unskip,\iSLAC}
\mbox{P. Stamer\unskip,\iRUTG}
\mbox{R. Steiner\unskip,\iADEL}
\mbox{H. Steiner\unskip,\iLBL}
\mbox{M.G. Strauss\unskip,\iMASS}
\mbox{D. Su\unskip,\iSLAC}
\mbox{F. Suekane\unskip,\iTOHO}
\mbox{A. Sugiyama\unskip,\iNAGO}
\mbox{S. Suzuki\unskip,\iNAGO}
\mbox{M. Swartz\unskip,\iSLAC}
\mbox{A. Szumilo\unskip,\iWASH}
\mbox{T. Takahashi\unskip,\iSLAC}
\mbox{F.E. Taylor\unskip,\iMIT}
\mbox{J. Thom\unskip,\iSLAC}
\mbox{E. Torrence\unskip,\iMIT}
\mbox{N. K. Toumbas\unskip,\iSLAC}
\mbox{A.I. Trandafir\unskip,\iMASS}
\mbox{J.D. Turk\unskip,\iYALE}
\mbox{T. Usher\unskip,\iSLAC}
\mbox{C. Vannini\unskip,\iPISA}
\mbox{J. Va'vra\unskip,\iSLAC}
\mbox{E. Vella\unskip,\iSLAC}
\mbox{J.P. Venuti\unskip,\iVAND}
\mbox{R. Verdier\unskip,\iMIT}
\mbox{P.G. Verdini\unskip,\iPISA}
\mbox{S.R. Wagner\unskip,\iSLAC}
\mbox{D. L. Wagner\unskip,\iCOLO}
\mbox{A.P. Waite\unskip,\iSLAC}
\mbox{Walston, S.\unskip,\iOREG}
\mbox{J.Wang\unskip,\iSLAC}
\mbox{C. Ward\unskip,\iBRUN}
\mbox{S.J. Watts\unskip,\iBRUN}
\mbox{A.W. Weidemann\unskip,\iTENN}
\mbox{E. R. Weiss\unskip,\iWASH}
\mbox{J.S. Whitaker\unskip,\iBU}
\mbox{S.L. White\unskip,\iTENN}
\mbox{F.J. Wickens\unskip,\iRAL}
\mbox{B. Williams\unskip,\iCOLO}
\mbox{D.C. Williams\unskip,\iMIT}
\mbox{S.H. Williams\unskip,\iSLAC}
\mbox{S. Willocq\unskip,\iSLAC}
\mbox{R.J. Wilson\unskip,\iCSU}
\mbox{W.J. Wisniewski\unskip,\iSLAC}
\mbox{J. L. Wittlin\unskip,\iMASS}
\mbox{M. Woods\unskip,\iSLAC}
\mbox{G.B. Word\unskip,\iVAND}
\mbox{T.R. Wright\unskip,\iWISC}
\mbox{J. Wyss\unskip,\iPADO}
\mbox{R.K. Yamamoto\unskip,\iMIT}
\mbox{J.M. Yamartino\unskip,\iMIT}
\mbox{X. Yang\unskip,\iOREG}
\mbox{J. Yashima\unskip,\iTOHO}
\mbox{S.J. Yellin\unskip,\iUCSB}
\mbox{C.C. Young\unskip,\iSLAC}
\mbox{H. Yuta\unskip,\iAOMORI}
\mbox{G. Zapalac\unskip,\iWISC}
\mbox{R.W. Zdarko\unskip,\iSLAC}
\mbox{J. Zhou\unskip.\iOREG}

\it
  \vskip \baselineskip                   
  \centerline{(The SLD Collaboration)}   
  \vskip \baselineskip        
  \baselineskip=.75\baselineskip   
\iADEL
  Adelphi University,
  South Avenue-   Garden City,NY 11530, \break
\iAOMORI
  Aomori University,
  2-3-1 Kohata, Aomori City, 030 Japan, \break
\iBOLO
  INFN Sezione di Bologna,
  Via Irnerio 46    I-40126 Bologna  (Italy), \break
\iBRUN
  Brunel University,
  Uxbridge, Middlesex - UB8 3PH United Kingdom, \break
\iBU
  Boston University,
  590 Commonwealth Ave. - Boston,MA 02215, \break
\iCINC
  University of Cincinnati,
  Cincinnati,OH 45221, \break
\iCOLO
  University of Colorado,
  Campus Box 390 - Boulder,CO 80309, \break
\iCOLU
  Columbia University,
  Nevis Laboratories  P.O.Box 137 - Irvington,NY 10533, \break
\iCSU
  Colorado State University,
  Ft. Collins,CO 80523, \break
\iFERR
  INFN Sezione di Ferrara,
  Via Paradiso,12 - I-44100 Ferrara (Italy), \break
\iFRAS
  Lab. Nazionali di Frascati,
  Casella Postale 13   I-00044 Frascati (Italy), \break
\iILLI
  University of Illinois,
  1110 West Green St.  Urbana,IL 61801, \break
\iLBL
  Lawrence Berkeley Laboratory,
  Dept.of Physics 50B-5211 University of California-  Berkeley,CA 94720, \break
\iLTU
  Louisiana Technical University,
  , \break
\iMASS
  University of Massachusetts,
  Amherst,MA 01003, \break
\iMISSI
  University of Mississippi,
  University,MS 38677, \break
\iMIT
  Massachusetts Institute of Technology,
  77 Massachussetts Avenue  Cambridge,MA 02139, \break
\iMOSCOW
  Moscow State University,
  Institute of Nuclear Physics  119899 Moscow  Russia, \break
\iNAGO
  Nagoya University,
  Nagoya 464 Japan, \break
\iOREG
  University of Oregon,
  Department of Physics  Eugene,OR 97403, \break
\iOXF
  Oxford University,
  Oxford, OX1 3RH, United Kingdom, \break
\iPADO
  Universita di Padova,
  Via F. Marzolo,8   I-35100 Padova (Italy), \break
\iPERU
  Universita di Perugia, Sezione INFN,
  Via A. Pascoli  I-06100 Perugia (Italy), \break
\iPISA
  INFN, Sezione di Pisa,
  Via Livornese,582/AS  Piero a Grado  I-56010 Pisa (Italy), \break
\iRAL
  Rutherford Appleton Laboratory,
  Chiton,Didcot - Oxon OX11 0QX United Kingdom, \break
\iRUTG
  Rutgers University,
  Serin Physics Labs  Piscataway,NJ 08855-0849, \break
\iSLAC
  Stanford Linear Accelerator Center,
  2575 Sand Hill Road  Menlo Park,CA 94025, \break
\iSOGA
  Sogang University,
  Ricci Hall  Seoul, Korea, \break
\iSOONG
  Soongsil University,
  Dongjakgu Sangdo 5 dong 1-1    Seoul, Korea 156-743, \break
\iTENN
  University of Tennessee,
  401 A.H. Nielsen Physics Blg.  -  Knoxville,Tennessee 37996-1200, \break
\iTOHO
  Tohoku University,
  Bubble Chamber Lab. - Aramaki - Sendai 980 (Japan), \break
\iUCSB
  U.C. Santa Barbara,
  3019 Broida Hall  Santa Barbara,CA 93106, \break
\iUCSC
  U.C. Santa Cruz,
  Santa Cruz,CA 95064, \break
\iVAND
  Vanderbilt University,
  Stevenson Center,Room 5333  P.O.Box 1807,Station B  Nashville,TN 37235,
\break
\iWASH
  University of Washington,
  Seattle,WA 98105, \break
\iWISC
  University of Wisconsin,
  1150 University Avenue  Madison,WS 53706, \break
\iYALE
  Yale University,
  5th Floor Gibbs Lab. - P.O.Box 208121 - New Haven,CT 06520-8121. \break

\rm
%

\end{center}

\vskip 1truecm

\end{document}